\documentclass{emulateapj}

\shorttitle{Strong Merger shock in A665}
\shortauthors{DASADIA ET AL.}

\def\chandra    {{\em Chandra}\/}

\begin{document}

\title{ A Strong Merger shock in Abell 665}

\author{
S. Dasadia\altaffilmark{1}, 
M. Sun\altaffilmark{1}, 
C. Sarazin\altaffilmark{2} 
A. Morandi\altaffilmark{1},
M. Markevitch\altaffilmark{3},
D. Wik\altaffilmark{3,4},
L. Feretti\altaffilmark{5},\\
G. Giovannini\altaffilmark{6},
F. Govoni\altaffilmark{7},
V. Vacca\altaffilmark{7}
}

\altaffiltext{1}{Physics Department, University of Alabama in Huntsville, Huntsville, AL 35816, USA; sbd0002@uah.edu; ms0071@uah.edu}
\altaffiltext{2}{Department of Astronomy, University of Virginia, Charlottesville, VA 22904, USA}
\altaffiltext{3}{Astrophysics Science Division, NASA/Goddard Space Flight Center, Greenbelt, MD 20771, USA}
\altaffiltext{4}{The Johns Hopkins University, Homewood Campus, Baltimore, MD 21218, USA}
\altaffiltext{5}{INAF - ORA Bologna, Via Gobetti 101, 40129 Bologna, Italy }
\altaffiltext{6}{Dipartimento di Astronomia, via Ranzani 1, 40127 Bologna, Italy}
\altaffiltext{7}{INAF - ORA Cagliari, Via della Scienza 5, I--09047 Selargius (CA), Italy}

\begin{abstract}

Deep (103 ks) \chandra\ observations of Abell 665 have revealed rich structures in this merging galaxy cluster, including a strong shock and two cold fronts. The newly discovered shock has a Mach number of $M$ = 3.0 $\pm$ 0.6, propagating in front of a cold disrupted cloud. This makes Abell~665 the second cluster where a strong merger shock of $M \approx$ 3 has been detected, after the Bullet cluster. The shock velocity from jump conditions is consistent with (2.7 $\pm$ 0.7) $\times$ 10$^3$ km sec$^{-1}$. The new data also reveal a prominent southern cold front, with potentially heated gas ahead of it. Abell 665 also hosts a giant radio halo. There is a hint of diffuse radio emission extending to the shock at the north, which needs to be examined with better radio data. This new strong shock provides a great opportunity to study the re-acceleration model with the X-ray and radio data combined.

\end{abstract}


\section{Introduction}

Large-scale structures such as galaxy clusters are natural outcomes of the hierarchical mergers of smaller subunits. Such mergers are among the most energetic events in the universe \citep[e.g.,][]{m1}. A large fraction of this energy is dissipated into the intracluster medium (ICM) through heating the ICM gas via shocks \citep[e.g.,][]{m2,s1}. Observationally, shocks can be seen as surface brightness edges in X-ray observations. These edges provide measurements of gas bulk velocities and can be used to study the micro transport processes in the ICM like thermal conduction and electron-ion equilibration \cite[e.g,][]{s2,t1,m2}. Shock detection is challenging, requiring favorable merger geometry. They are therefore rare and only a handful of shocks have been detected so far. Typically shocks in merging galaxy clusters have relatively low Mach numbers ($M \leq 3$) and only two clusters have, $M$ $\geq 2.5$ i.e., A521 \citep{b1} and the Bullet cluster \citep{m3}.

As the richest cluster classified by \cite{a3}, Abell 665 (A665 from now) has been studied extensively at many wavelengths. \cite{g3} in their optical study noticed an elongation in the galaxy distribution suggesting the unrelaxed state of the cluster. On the other hand, \citet[hereafter GHB00]{g2} studied the dynamics of the cluster and only found weak evidence for substructures in radial velocities of galaxies. The cluster hosts a giant radio halo first observed by \cite{m6}. \citet{v1} examined the radio halo emission in detail with deep {\em VLA} observations and constrained the magnetic field distribution in the cluster. Previous X-ray observations revealed complex substructures including possible detection of a shock in front of a core which appeared to correlate with the radio emission \citep[e.g.,][]{m8,g4}. 

To further investigate, we obtained deep \chandra\ observations in 2012. In this letter, we present the detection of a strong shock from new observations. Throughout the letter, $H_{0}$ = 70 km s$^{-1}$ Mpc$^{-1}$, $\Omega_{M}$ = 0.3, $\Omega_{\Lambda}$ = 0.7. At $z = 0.182$ (GHB00), 1$''$ =  3.05 kpc. All error bars reported show 68$\%$ confidence interval.

\section{Data Analysis}

A665 was observed by Advanced CCD Imaging Spectrometer (ACIS) in the Very Faint mode for a total exposure of $\sim$ 125 ks. All observations were taken with the ACIS-I. \chandra\ Interactive Analysis of Observations v4.7 and calibration database v4.6.7 from \chandra\ X-ray Center were used for data reduction. We followed the data analysis described in \cite{d1}. The clean exposures are: 32.7 ks for ObsID 12286, 48.5 ks for ObsID 13201 and 22.2 ks for ObsID 3586. Point sources were excluded from spatial and spectral analysis. 

The $N_H$ value towards the cluster, 5.0 $\times$ 10$^{20}$ cm$^{-2}$, was determined from the spectrum of a large region of the cluster excluding the core. This is the same as the value predicted from the empirical relation by \cite{w1}, while the HI column density towards the cluster is 4.2 $\times$ 10$^{20}$ cm$^{-2}$ \citep[][]{k1}. Reducing $N_H$ to the HI value only causes 7 - 10\% changes on best-fit parameters. We followed the same background method by \cite{s5}. Spectra were fitted with XSPEC version 12.8 \citep{a1}. This work used AtomDB 2.0.2. Throughout the analysis, we adopt the solar abundance table by \cite{a2}.

\section{Results}

Figure 1A shows the exposure corrected image of the cluster produced by combining all three observations in the 0.7 - 2.0 keV energy band. The image reveals disturbed gas with multiple discontinuities caused by the merger. The X-ray morphology suggests a violent merger where the two subclusters have recently passed through each other roughly along the north-south direction \citep[e.g.,][]{m8}. In this letter, we call the one merging from the north as the subcluster and the other one as the primary cluster. The X-ray peak is elongated in the north-west (NW) to south-east (SE) direction with a bright edge in the south (C2). This X-ray brightest region, as the core of the infalling subcluster, is relatively cool. The ram pressure generated by the motion of the subcluster is stripping materials from the bright core. There is a tail of emission that can be seen to the north over a distance of $\sim$ 300 kpc. No surface brightness peak associated with the primary cluster core is observed. It is likely the primary cluster's core has already been disrupted by the merger. We notice two additional edges (C1 \& S1) in the north. The inner edge (C1) marks a boundary between the remnant core of the primary cluster and the ICM of the subcluster. There is a second, outer surface brightness edge (S1) $\sim$ 600 kpc ahead of the inner edge. We confirm that this is a shock which separates shock heated gas from the cluster outskirts. In this letter, we mainly focus on these two northern edges.

\subsection{Spatial Analysis}

Figure 1B shows the surface brightness profile in the northern sector N. The profile changes its slope at two places, C1 (the cold front) and S1 (the shock front). Regions I-III (the remnant core, the postshock region, and the preshock region) can then be defined.
The gas density distribution in regions I \& II was derived by fitting corresponding surface brightness profiles to a model, where the X-ray emissivity ($\epsilon$) and radius ($r$)  are related by a powerlaw, $\epsilon \propto r^{-p}$ within each region assuming an ellipsoidal geometry as discussed in \cite{k2}. The model for a region with two edges was fitted. The best-fit powerlaw index ($p$) in region I \& II are 0.60 $\pm$ 0.01 and 1.1 $\pm$ 0.1 respectively. Using the powerlaw index, we reconstructed the intrinsic emissivity distribution to obtain the corresponding density distribution, $n_e(r)$ = $[\epsilon(r)/\Lambda(T_e, Z)]^{1/2}$, where $\Lambda(T_e, Z)$ is the X-ray emissivity function which depends on electron temperature $T_e$ and abundance $Z$. The surface brightness in the preshock region (Region III) was fitted to a powerlaw density model, $n_e(r)$ = $n_{e0}  r^{-\alpha}$. The model provides a good fit to the data and $\alpha$ = 2.4 $\pm$ 0.3. 

Figure 1D shows the best-fit density model in each region. The density jumps across the shock by a factor of $\rho_{2}$/$\rho_{1}$ = 3.2 $\pm$ 0.3, where suffix 1 and 2 describe quantity before and after the shock. We apply Rankine-Hugoniot jump conditions to determine the Mach number $M$ = 3.4 ($>$ 2.7 at 90\%) for a monoatomic gas ($\gamma$ = 5/3) \citep[e.g.,][]{l1}. The models used to fit surface brightness profiles approximate the shape of edges to be elliptical. We have to make assumptions for the curvature of the isodensity surfaces along the line of sight, which introduces a systematic uncertainty. For example, if we assume that the shock surface has a curvature along the line of sight similar to that in the plane of the sky ($r\approx 2$ Mpc instead of the distance to the cluster center, 1.4 Mpc, that was used for the above fit), while the preshock gas is still centered on the cluster center, the shock density jump is reduced to factor 2.6, and the corresponding Mach number to 2.5.

Sector S (Figure 1A) was used to investigate the southern cold front (C2). Figure 1E includes the 0.7 - 2.0 keV surface brightness profile across C2. The slope change at $\sim$ 170 kpc corresponds to a density jump by a factor of $\sim$ 2.1. The profile shows no additional discontinuity ahead of C2.

\subsection{Spectral Analysis}

Sector N was divided into four smaller regions (I, IIa, IIb, III) in the spectral analysis. The pre-shock region (region III) also includes an extended source with unknown nature (Ear2 in $\S$4.1), which was excluded in the spectral analysis. For each region, the spectra from individual observations were fitted simultaneously to an absorbed single temperature thermal emission model APEC \citep[e.g.,][]{s3}, after accounting for the background (see Dasadia et al. 2016). The spectra were fitted in the 0.5 - 7.0 keV energy band and the best-fit parameters were obtained by minimizing C-statistics defined in XSPEC.

The best-fit projected (black) and deprojected (red) temperatures are shown in Figure 1C. A deprojection correction was applied by estimating the projection contribution of outer layers onto the inner ones. The layers were assumed to be spherical in the shape, which is a reasonable assumption for wedge shaped regions across fronts. The gas temperature beyond region III, 2.5 keV, was estimated from the azimuthally averaged temperature profile of the cluster.  For the deprojection analysis, we combined regions II a \& b. The best-fit density model was then multiplied by the corresponding temperature to obtain the electron pressure distribution in Figure 1D.

Region I contains gas at a temperature of 8.2$_{-0.6}^{+0.8}$ keV and abundance of 0.3 $\pm$ 0.1 $Z_{\bigodot}$. Across the inner edge (C1), the surface brightness drop is accompanied by an increase in the temperature. Region II contains the hottest gas in the cluster with an average temperature of $\sim$ 11 keV. The best-fit temperature in region IIb is 11.3$_{-1.8}^{+2.2}$ keV. The abundance in region IIb is poorly constrained. Thus, we fixed the abundance to 0.2 Z$_{\bigodot}$. The best-fit temperature changes by less than 5\% for $Z$ = 0.1 $Z_{\bigodot}$ - 0.4 $Z_{\bigodot}$. We combine regions II a \& b to obtain postshock temperature of $T_{shock}$ = 11.0$_{-0.8}^{+1.4}$ keV. The preshock temperature from region III is $T_{preshock}$ = 3.2 $\pm$ 1.1 keV. Thus, the temperature jumps across the shock front by a factor of 3.8 $\pm$ 1.3, which corresponds to a Mach number $M_{proj}$ = 3.0 $\pm$ 0.6.  The Mach number after applying deprojection correction is $M_{deproj}$ = 3.2 $\pm$ 0.7. This is consistent with the Mach number derived from the density jump. The preshock sound speed is $c_s$ = (9.1 $\pm$ 0.1) $\times$ 10$^2$ km sec$^{-1}$, giving a shock speed of $v_{shock} = M_{proj} c_s$ = (2.7 $\pm$ 0.7) $\times$ 10$^3$ km sec$^{-1}$. 

The velocity of the cold disrupted cloud can also be estimated from the pressure ratio between just inside the cold cloud and the free stream, assuming a uniform flow of gas around a blunt body. While this is a good approximation for simple mergers, it may not be true for complex mergers like in A665. For example, the shock is separated from the cold front by $\sim$ 600 kpc, which is much larger than expected for a steady-state situation. Also, in models of cluster mergers after the first core passage, the subcluster core is slowed by gravity and drag and falls back into the central region of the merged cluster, while the shock speeds up towards the cluster outskirt by following a negative pressure gradient.

\section{Discussion}

\subsection{Merger Dynamics}
 
The obvious interpretation of the data is that the cluster is experiencing a two-body merger. A study on cluster dynamics of A665 done by GHB00 concluded that the velocity distribution is similar to a fairly relaxed and massive cluster and detected ``only marginal evidence for substructure and non-Gaussianity in the velocity distribution". This may imply that the merger is occurring nearly in the plane of sky. The $N$-body simulation by GHB00 suggests that the observed velocity dispersion is consistent with a merger between two similar sized subclusters caught in the middle.  The X-ray data reveal one core while the other core may have been disrupted by the merger. 
 
Elongation of the X-ray peak suggests that the merger axis is close to N - S. The infalling subcluster core is moving south and is being stripped by ram pressure. The surface brightness edge ahead the remnant core is a cold front (C2) formed by the head-on merger. Previously, \citet[O09 hereafter]{o3} argued that the southern edge (C2) in Abell 665 is more likely a shock. O09 fitted the spectrum for the cold front using a region which extended beyond the cold front. In any case, the errors on the spectral parameters in O09 allow for either a shock or a cold front. From our new data, we find that the temperature increases by a factor of $\sim$ 1.3 (from $7.3 \pm 0.4$ keV to $9.3 \pm 0.7$ keV) across C2 which confirms it is a cold front (Figure 1F). The surface brightness distribution derived for sector S (Figure 1E) shows no slope change outside of C2. However, the relatively high temperature ahead of C2 may imply a shock heated region there \citep[as suggested in][]{m8}, although the current \chandra\ data do not allow a good constraint on the temperature beyond sector S.

We derived the abundance map of the cluster, which is nearly flat at $\sim$ 0.2 $Z_{\bigodot}$. The core of the subcluster shows no abundance enhancement. This is in contrast to merging clusters with strong cool cores. For example, in A2146, the abundance at the subcluster core ($\sim$ 0.9 $Z_{\bigodot}$) is much higher than elsewhere at $\sim$ 0.4 $Z_{\bigodot}$ \citep[][]{r1}. Similarly, in $RX J0334.2-0111$ there is a clear abundance difference between two merger components \citep[][]{d1}. On the other hand, A665's core is at most a weak cool core, with a central entropy of $\sim$ 135 keV cm$^{2}$ \citep{cav09}.

The more diffuse merger component can be seen in the north with an upstream cold front (C1). C1 has a large radius of curvature, $R_{cf}$ $\sim$ 750 kpc. The edge appears sharper in the east and can be traced over a distance of $\sim$ 850 kpc. The distance between the stagnation point and the closest point on the shock (S1), also known as ``stand-off" distance $d_s$, is $\sim$ 600 kpc. This is the largest among all merging clusters e.g., in the case of the Bullet cluster, stand-off distance is $\sim$ 138 kpc \citep[][]{m3} while in A754 it is approximately $\sim$ 160 kpc \citep[][]{m5}. The shock has an almost flat shape and is visible over a distance of $\sim$ 1.2 Mpc. This is similar to the shock length $\sim$ 1.1 Mpc in the merging cluster A520 \citep[][]{m4}. However, in A520 the stand-off distance is only 50 kpc. 

Two substructures (Ear1 \& Ear2) on the north of the cluster have been observed (Figure 1A). Ear1 is located $\sim$ 1.5 Mpc from the cluster center and the faint galaxies within it have no redshift information. If Ear1 is an isolated system, its redshift range can be estimated from the $L - T$ relation \citep[e.g.,][]{s4}. The allowed $z$ range is 0.5 - 0.9 with a system temperature of $\sim$ 4 keV. Ear2 is located $\sim$ 330 kpc north of S1 and contains two galaxies at the cluster redshift. The allowed $z$ range of Ear2, from the $L - T$ relation, is 0.1 - 0.3. Thus, Ear2 is likely another subcluster associated with A665. Ear2 was excluded in our analysis for the pre-shock region.

 \subsection{Diffuse Radio Emission}
 
  A665 contains a giant radio halo (RH) which is about 1.8 Mpc in size \citep[][]{g1}. Recent VLA observations show that the RH is elongated in the SE-NW direction. It appears that the radio emission is asymmetric with respect to the cluster center, being brighter and more extended towards NW \cite[e.g.,][]{f2,v1}. The spectral index map of the radio halo produced by \cite{f2} shows that the radio spectrum near the X-ray shock is flatter than that in the southern region, which may be consistent with the recent shock acceleration observed. There is some radio emission close to S1. After removing the contribution from several point sources using multi-resolution radio data, the flux associated with the diffuse region near the X-ray shock is $1.9 \pm 0.2$ mJy, although the source subtraction is uncertain and the presence of this emission needs confirmation with deeper observations.
  
X-ray shock fronts often exhibit a corresponding sharp feature in the radio synchrotron emission \citep{m9}. It is interesting to note that the relatively strong ($M \approx 3$) shock front that we observe in A665 does not exhibit a prominent radio relic, as do some other shocks with similar or lower Mach numbers, such as A521 \citep[$M \approx 2.5$;][]{g5,b1}, A754 \citep[$M\approx 1.6$;][]{k3,m5}, A3667 \citep[$M \approx 2$;][]{f3} and the eastern shock in the Bullet cluster \citep[$M\approx 2.5$;][]{s6}. It also does not exhibit (at least with the present radio sensitivity) an abrupt edge of the giant halo at the shock, as observed, e.g., at the western shock in the Bullet cluster \citep[$M \approx 3$;][]{ m3,s7}, A520 \citep[$M\approx 2.3$;][]{m4} and Coma \citep[$M\approx 2$;][]{b2}. 
  
 A665 appears similar to A2034 \citep[$M\approx 1.6$;][]{o2} in that they both exhibit only hints of diffuse radio emission in the post-shock region. Some other shocks clearly visible in X-rays do not show any radio emission --- e.g., the two $M \approx 1.6 - 2$ fronts in A2146 \citep{r1}. Radio relics are believed to be produced by diffusive acceleration of ultrarelativistic electrons on shock fronts \citep[most likely starting with aged relativistic electron population rather than thermal electrons, e.g.,][]{m5, b3}. But the above examples show that it occurs on some shocks but not others, apparently unrelated to the Mach number. This may be explained by different density of seed electrons available for re-acceleration at the shock locations --- relics occur only when a shock front crosses a cloud with an excess of such electrons, as suggested in \cite{s6}. In other locations, shocks either do not produce detectable radio emission, or produce only a faint edge of a giant halo, within which the merger turbulence takes over and provides re-acceleration needed to sustain the halo.
 
\section{Summary}

Deep \chandra\ observations of A665 have revealed complex substructures, including the detection of a new strong shock with a Mach number of 3.0 $\pm$ 0.6 and at least two cold fronts. The shock propagates in front of the cooler gas, apparently a remnant of a dense core which may have been disrupted by the merger. The shock front can be traced to $\sim$ 1.2 Mpc in length and has a velocity of 2.7 $\pm$ 0.7 $\times$ 10$^3$ km sec$^{-1}$. The system exhibits a unique merger geometry, e.g., the distance between the shock and the stagnation point of a cold front, $\sim$ 600 kpc, is largest among all merger shocks. This provides an excellent opportunity to study shock heating mechanism, although this will require deeper X-ray data. We also rule out the prominent southern edge as a shock, as suggested previously; it is in fact a cold front. 

The cluster hosts a giant ($\sim$ 1.8 Mpc in size) radio halo which appears to extend south of cold front (C2). There is a radio extension towards the northern shock. However, given the contamination of discrete sources located near the shock location, the spatial correlation between the X-ray shock and the diffuse radio emission must be taken with caution. Mass distribution in A665, constrained from lensing, would be important to understand the dynamics in A665. The early result by \cite{d2} suggested an offset of the X-ray peak from the galaxy distribution, which needs to be better studied. 

\acknowledgments

The support for this work was provided by the National Aeronautics and Space Administration (NASA) through \chandra\ awards GO2-13160A and GO2-13102A issued by the \chandra\ X-ray Observatory Center which is operated by the Smithsonian Astrophysical Observatory for and on behalf of the NASA under contract NAS8-03060.

\begin{figure*}
\vspace{-2.5 cm}
\hspace{-0.7 cm}
\includegraphics[width=1.05\textwidth]{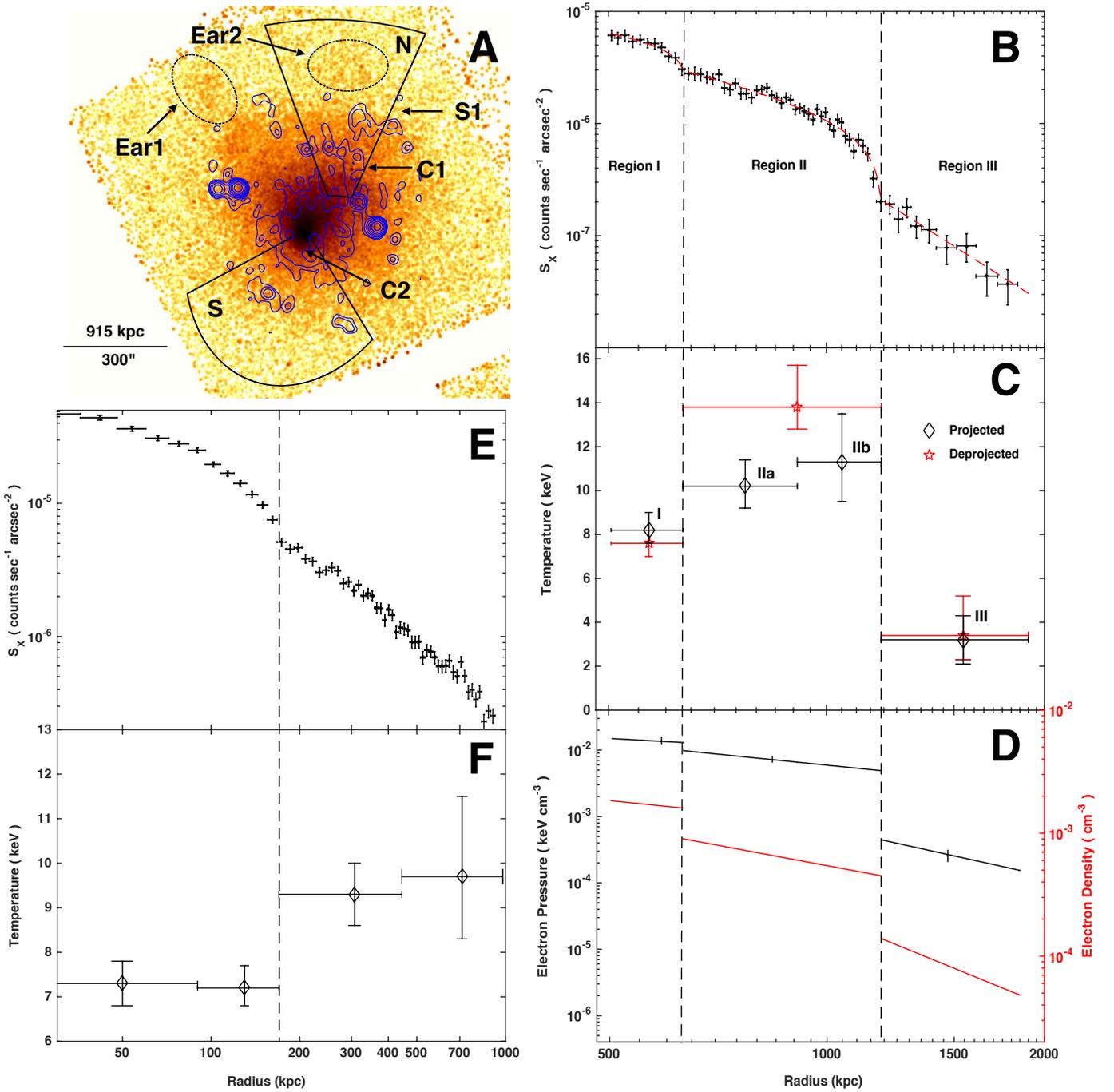}
\vspace{-3 cm}
\caption{A: The background subtracted, exposure corrected \chandra\ image of A665 in the 0.7 - 2.0 keV energy band, smoothed by a Gaussian with $\sigma = 5.9''$. Point sources are removed. The arrows show the southern cold front (C2), the northern cold front (C1) and the shock front (S1). Two diffuse sources are marked by Ear1 and Ear2. Sectors N and S used to study surface brightness edges are also shown. The $VLA$ 1.4 GHz contours (blue) from \citet{v1} are also overlaid (contour levels from 135 $\mu$Jy/beam to 17 mJy/beam, spaced by a factor of 2).
% The correlation between X-ray shock and radio emission is marginal due to nearby point sources. 
B: The 0.7-2.0 keV surface brightness profile in sector N. Three distinct regions are marked: inside cold front (Region I), post shock (Region II) and pre shock (Region III). The two vertical dashed lines show the location of the cold front (left) and the shock front (right). The red dashed line shows the surface brightness profile from the best-fit density model. C: The projected (black) and deprojected (red) temperature values in regions (I, IIa, IIb, \& III) across northern edges C1 \& S1. D: The corresponding model density (red) and pressure (black) profiles. Error bars on pressure show 1$\sigma$ confidence interval. E: The 0.7-2.0 keV surface brightness profile in sector S. The location of C1 is marked by a vertical dashed line. The profile show no discontinuity ahead of C1. F: Projected temperature profile across C1 that suggests C1 as a cold front.} 

\end{figure*}

\end{document}